\documentclass{bb}  
\usepackage{graphicx}
\usepackage{txfonts}
\begin{document}

\title{A search for RCB stars in globular clusters}
   
\titlerunning{A search for RCB stars in globular clusters}

\author{Scarlett-Rose Boiardi\inst{1,2},
S\'ebastien Roger\inst{1,3}
\and Emmanuel Davoust\inst{1}
}
\institute{
IRAP, Universit\'e de Toulouse, CNRS, 
 14 Avenue Edouard Belin, F-31400 Toulouse, France
\and
The University of Warwick, Coventry, CV4 8UW, United Kingdom
\and
Loughborough University, Loughborough, Leicestershire, LE11 3TU, United Kingdom
}

\date{ }

\abstract
{There are only about 65 R Coronae Borealis stars known in our Galaxy, and none
in globular clusters. As these stars are thought to result from the merger of two
white dwarfs, one would expect the higher stellar density of globular clusters to
favor their formation.

We have searched for such stars in Galactic globular clusters,
as their presence in a specific category of clusters might provide more clues as to their
formation.

We selected from the WISE all-Sky source catalog all the stars within the tidal radius
of the 150 globular clusters within 50 kpc, which is the distance to which RCB stars are
detectable by WISE. The total number of stars selected in this way was 635989.
We then successively applied the eight selection criteria of 
Tisserand (2012) satisfied by RCB stars to the dereddened photometric WISE and 2MASS data.

Only three stars satisfying the conditions were found in the field of three globular clusters.
The star in the field of Liller 1 is most probably a protostar. For the two
other candidates, the absence of photometry in the visible range did not allow us to
establish their nature with certainty.
We further identified one dust-enshrouded star that only satisfied the first selection criteria,
and used DUSTY to determine that it is a star of temperature 4800K enshrouded in a 
dusty envelope with a temperature 300 K and an opacity in the visible of 0.59.
It is probably an Xray binary star with a dusty accretion disk.

We found no RCB stars truly belonging to a globular cluster, thus providing a constraint
on their formation mechanism.
}
    

\maketitle

%________________________________________________________________
\section{Introduction}
\label{intro}

R Coronae Borealis stars are a very small group of hydrogen-deficient and carbon-rich supergiants. 
About 65 are known in the Galaxy and 25 in the Magellanic Clouds. In our Galaxy, they are
mostly confined to low Galactic latitudes, and are thus thought to be part of the bulge 
population. Two scenarios have been proposed to explain their existence, either the double
degenerate merger of two white dwarfs, or the final helium shell flash of the central star
in a planetary nebula (see Clayton 2012 for a review).

The expected number of RCB stars in the Galaxy can be as large as 500 (Tisserand 2012) or even
5000 (Clayton 2012).  It thus seems feasible to increase the number of known RCB stars 
significantly, in order to better understand their evolutionary status. 
Since these stars are carbon-rich, they should be more easily detectable from their
infrared properties.  Tisserand (2012) recently proposed eight criteria for identifying
RCB stars from their mid-infrared WISE colors.

Our own interest lies with the stellar populations in globular clusters (Sharina \& Davoust 2009,
Sharina et al. 2010), and we recently discovered a CH star in the globular cluster NGC 6426
(Sharina et al. 2012). Since the presently more likely formation scenario for RCB stars 
seems to be a merger of white dwarfs, the dense environment of globular clusters should be
ideal for their formation. We thus decided to take advantage of the recently available
WISE all-sky source catalogue to search for RCB stars in globular clusters, using the selection
criteria of Tisserand (2012). An advantage of searching for such stars in globular clusters 
is that the distance and reddening are known {\it a priori}.

%__________________________________________________________________
\section{Selection of the candidate RCB stars}
\label{selection}

We selected the target globular clusters from the catalogue of Harris (1996), 2010 edition.
Since RCB stars are detectable in the WISE bands out to 50 kpc (Tisserand 2012), we removed
the more distant clusters.  A total of 150 clusters were thus retained.
We then extracted the WISE and 2MASS photometry from the WISE all-sky source catalogue,
selecting all the stars within the tidal radius of each cluster, which was computed from
the catalogue of Harris (1996), 2010 edition. This produced a catalogue of 635989 stars.

The magnitudes were dereddened using the standard procedure (Cardelli et al. 1989).
For the four WISE bands W1, W2, W3 and W4, the total extinctions
$R_\lambda$ are 0.158, 0.093, 0.087 and 0.056, respectively (Bilir et al. 2011).
The distance and foreground reddening for each globular cluster were taken from Harris (1996, 2010 edition).

The first step was to apply the first four selection criterion of Tisserand (2012) to the
stars. This left 2581 stars in 84 clusters. We then extracted K-band images of all the stars
from the Interactive 2MASS Image Service and inspected them for possible problems
(artefact, nearby star, etc.) that could affect the photometry of each star. 
We considered this lengthier
procedure safer than relying on the WISE and 2MASS flags. A total of 776 stars and two clusters
were thus removed.

After applying the last four selection criteria of Tisserand (2012), we were left with only
three candidate RCB stars in three globular clusters.  Their coordinates and photometry are given in Table~\ref{data}.  
None of the three stars is in the list of RCB candidates of Tisserand (2012, his Table 5).
It is rather disturbing that all three candidates are situated at a very low galactic latitude ($b \leq 0.16$ degrees);
this suggests that they are all field stars that happen to be in the field of a globular cluster and thus
that the adopted distances and reddenings might be incorrect. 

\begin{table*}
%\begin{minipage}{\columnwidth}
\caption{Magnitudes and distance to the cluster center (in units of the tidal radius) for the three candidate RCB stars
}
\label{data}
\centering
\begin{tabular}{lrrrrrrrrlr}
\hline
Name&I&J&H&K&W1&W2&W3&W4&cluster&$r/r_t$\\
\hline
J173319.71-332320.1&21.82&13.945&11.753&10.167&9.258&8.653&7.832&7.858&Liller 1&0.08\\
J180844.18-194928.4&     &17.979&15.788&12.876&10.770&10.068&7.941&7.211&2MS GC01&0.70\\
J184931.40-013314.7&     &16.060&13.909&12.232&10.972&10.560&8.378&7.771&Glimpse 01&0.79\\
\hline
\end{tabular}
%\end{minipage}
\end{table*}

%__________________________________________________________________
\section{Analysis of the three candidates}
\label{analysis}

\begin{figure}
\centering
\includegraphics[angle=-90,width=8.5 true cm]{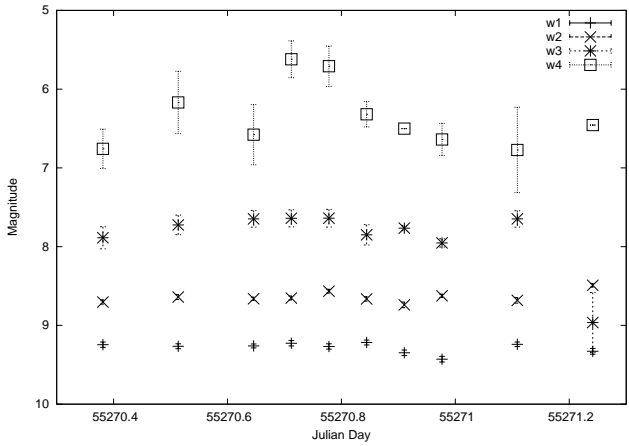}
\caption{WISE magnitudes of the candidate RCB star in Liller 1 vs Julian Day.
The dusty shell became slightly brighter in the redder two bands halfway through the observations. 
}
\label{fig_wisevar}
\end{figure}

The candidate RCB star in the field of \object{Liller 1} has ten observations in the WISE single-exposure catalogue. 
We plotted the four WISE magnitudes of the star against Julian Day on Fig.~\ref{fig_wisevar},
which shows that the dusty shell became slightly brighter in the redder two bands halfway through the observations, with a hint of 
oscillatory behavior.

\begin{figure}
\centering
\includegraphics[width=8.5 true cm]{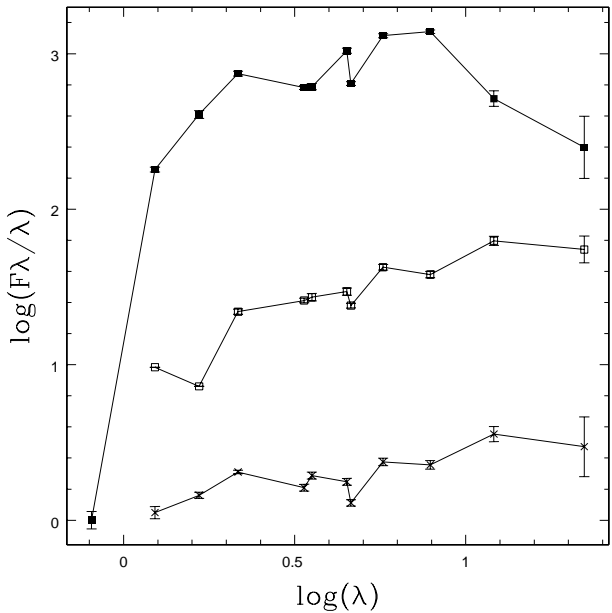}
\caption{Spectral energy distribution of the three candidate RCB stars.
From top to bottom : in Liller 1, 2MS-GC01 and GLIMPSE01.
The fluxes in the four IRAC bands are also plotted.
The flux units are arbitrary.
}
\label{fig_3sed}
\end{figure}

An infrared color-magnitude diagram of \object{Liller 1} has been obtained by Valenti et al. (2010).
If we compare the magnitudes of the star given by Valenti et al. (2010), which date from 30 July 2007,
and those given by 2MASS, which were obtained on 13 August 1998, the star did not vary in H and K and
became 0.5mag brighter in J between 1198 and 2007. The K-mag given by 
DENIS\footnote{http://cdsweb.u-strasbg.fr/denis.html}, 
which dates from an earlier epoch than 2005, is comparable to the two other values.

We obtained images in V and I of \object{Liller 1} from the ESO archives, reduced them
in a standard way and applied sextractor (Bertin \& Arnouts 1996) to the images.
The zero-point of the I-band images was calibrated with photometry of 14 stars from 
DENIS.
Those of the V-band were calibrated by comparing our color-magnitude diagram
with that of Ortolani et al. (1996). Since the latter did not publish any catalogue, the
uncertainty on the zero-point in V is rather large, and estimated at 0.2 mag.
The derived magnitude in I of our star is $I = 21.82$, and it was not visible at all
on the V-band image. Since one expects a V-band apparent magnitude in the range 17.6 -- 19.6,
which is easily reached in the V-image,
this is not an RCB star if it is at the distance of \object{Liller 1}. In fact, the
spectral energy distribution, shown on Fig.~\ref{fig_3sed}, suggests that this is a protostar 
(see Fischer et al. 2012 for such a spectrum).

We were not able to find images or photometry in the visible range of the two other stars.
Their spectral energy distribution, also shown on Fig.~\ref{fig_3sed}, does not suggest that
they could be RCB stars.

%__________________________________________________________________
\section{Search for dust-enshrouded stars}
\label{four}

We returned to the sample of 1815 stars satisfying the first four criteria of Tisserand (2012) to search for
dust-enshrouded stars. We inspected the spectral energy distribution of all the stars, and found only 21
which had the shape expected from a hot star ($T_\star>$ 4000) with a warm dusty envelope ($800 > T_{dust} > 200$ K).
Out of these, only one, J174104.68-534245.4 in the field of NGC 6397 was not within a few degrees of the Galactic plane. 

Photometry in the visible and infrared ranges and proper motions for the star in the field of NGC 6397
were obtained using Vizier. It has non-zero proper motions, and thus cannot belong to the cluster.
It has also been detected by Chandra, and is thus probably an X-ray binary star with a dusty torus.

\begin{figure}
\centering
\includegraphics[width=8.5 true cm]{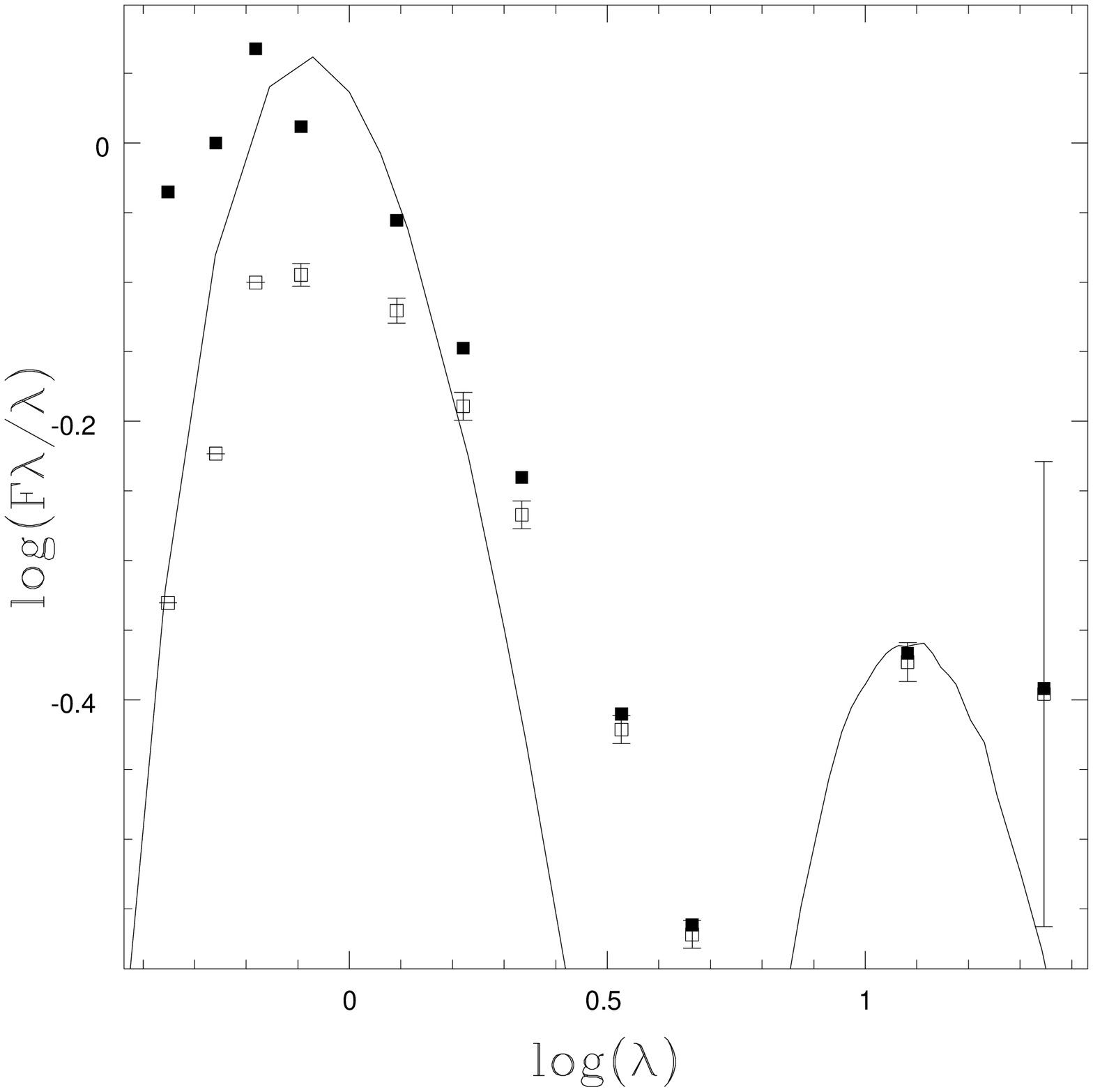}
\caption{Spectral energy distribution of the dust-enshrouded star
in the field of NGC 6397.
The solid squares are the fluxes corrected for an extinction of 0.18,
while the open squares are the uncorrected fluxes with error bars.
The solid curves show the model spectral energy distribution expected for a dust-enshrouded star.
The monochromatic fluxes are normalized to the observed flux in the B band.
}
\label{fig_sed6397}
\end{figure}

We fit a model composed of a radiation point-source and a dusty shell to the spectral energy distribution of the star,
using the publicly available software DUSTY (Nenkova et al. 2000). We assumed a central source of
temperature 4800 K and a shell of amorphous carbon grains with a standard MRN
distribution of sizes (Mathis et al. 1977), visual optical depth 0.58 and temperature 300 K.
The fit is shown on Fig.~\ref{fig_sed6397}. Since we do not know the exact distance to the star,
we plotted the fluxes corrected and uncorrected for extinction.

%__________________________________________________________________
\section{Conclusion}
\label{conc}

The negative result of our search for RCB stars in Galactic globular clusters indicates that
such stars are very rare, if not completely absent, in these old stellar systems. 
This may be because the RCB phenomenon has a very short lifetime, is a rare event in the
evolution of stars, or that it occurs in stars younger than five to seven Gyr, which is the
age range of the youngest clusters explored in this project. 

\begin{acknowledgements}
This report summarizes the results of a summer internship.
It made use of the databases Simbad and Vizier, operated at CDS, Strasbourg, France.
It made use of data products from the Wide-field Infrared Survey Explorer, 
which is a joint project of the University of California, Los Angeles, and the Jet Propulsion 
Laboratory/California Institute of Technology, funded by the National Aeronautics and Space 
Administration.
It made use of data products from the Two Micron All Sky Survey, which is a joint project 
of the University of Massachusetts and the Infrared Processing and Analysis Centre, California 
Institute of Technology, funded by the National Aeronautics and Space Administration and the 
National Science Foundation. 
It also made use of the DENIS database. 
The DENIS project has been partly funded by the SCIENCE and the HCM plans of the European Commission 
under grants CT920791 and CT940627. It is supported by INSU, MEN and CNRS in France, by the State of 
Baden-W\"urttemberg in Germany, by DGICYT in Spain, by CNR in Italy, by FFwFBWF in Austria, by FAPESP in 
Brazil, by OTKA grants F-4239 and F-013990 in Hungary, and by the ESO C\&EE grant A-04-046.
 
\end{acknowledgements}

\end{document}